\documentclass{desyproc}
\newcommand{\hefour}{\ensuremath{{}^{4}{\rm He}}}
\newcommand{\hethree}{\ensuremath{{}^{3}{\rm He}}}

\newcommand{\gag}{\ensuremath{g_{{\rm a}\gamma}}}
\newcommand{\ma}{\ensuremath{{m_{\rm a}}}}

\begin{document}
\title{The CAST experiment: status and perspectives}

\author{{\slshape Esther Ferrer Ribas on behalf of the CAST Collaboration\\[1ex]
IRFU, CEA/Saclay, 91190 Gif sur Yvette, France\\}}

\contribID{lindner\_axel}

\desyproc{DESY-PROC-2009-05}
\acronym{Patras 2009} 
\doi  

\maketitle

\begin{abstract}
The status of the solar axion search with the CERN Axion Solar Telescope (CAST) will be discussed. Results from the first part of CAST phase II where the magnet bores were filled with \hefour~gas at variable pressure in order to scan \ma~up to 0.4~eV will be presented. From the absence of excess X-rays when the magnet was pointing to the Sun, we set a typical upper limit on the axion-photon coupling of $\gag \lesssim ~2.17 \times$10$^{-10}$GeV$^{-1}$ at 95\% CL for \ma$<0.4$~eV, the exact result depending on the pressure setting. Our search for axions with masses up to about 1.2~ eV using \hethree~as a buffer gas is, since last year, in progress in the second part of CAST phase II. Expectations for sensibilities will be given. Near future perspectives as well as more long term options for a new helioscope experiment will be evoked.
\end{abstract}

\section{Introduction}

The CAST (Cern Axion Solar Telescope) experiment is using a
decommissioned LHC dipole magnet to convert solar axions into
detectable x-ray photons. Axions are light pseudoscalar particles
that arise in the context of the Peccei-Quinn\cite{PecceiQuinn}
solution to the strong CP problem and can be Dark Matter
candidates\cite{Sikivie}. Stars could produce axions via the
Primakoff conversion of the plasma photons. The CAST experiment
is pointing at our closest star, the Sun, aiming to detect solar axions. The
detection principle is based on the coupling of an incoming axion
to a virtual photon provided by the transverse field of an intense
dipole magnet, being transformed into a real, detectable photon
that carries the energy and the momentum of the original axion.
The axion to photon conversion probability is proportional to the
square of the transverse field of the magnet and to the active
length of the magnet. Using an LHC magnet ($9\;$T and $9.26\;$m long)
improves the sensitivity by a factor 100 compared to previous
experiments. 
The CAST experiment has been taking data since 2003 providing the most restrictive
limits on the axion-photon coupling~\cite{Zio05,And07} for masses 
$\ma\lesssim 0.02\;$eV. At this mass the sensitivity is degraded due to coherence loss. 
In order to restore coherence, the magnet can be filled with a buffer 
gas providing an effective mass to the photon\cite{vanBibber:1988ge}. By changing the pressure 
of the buffer gas in steps, one can scan an entire range of axion mass values.  
At the end of 2005 the CAST experiment started such a program, entering its phase II 
by filling the magnet bore with He gas. From 2005 to 2007, the magnet bore was filled with 
$^{4}{\rm He}$ gas extending our sensitivity to masses up to $0.4\;$eV, final results will be presented here. From March 2008 onwards the magnet bore has been filled with $^{3}{\rm He}$ and  
the sensitivity should be increased to sensivities up to $\ma\lesssim1.2\;$eV by the end 
of the $^{3}{\rm He}$ run in 2010. 

\section{The CAST experimental set-up: recent upgrades}

The CAST set up has been described elsewhere~\cite{Zio05,Zio99}. From 2002 to 2006
three X-ray detectors were mounted on the two sides of the magnet: a conventional TPC\cite{Aut07} 
covering both magnet bores looking for sunset axions; in the sunrise side one of the bores was covered by a Micromegas detector\cite{Abb07} and in the other bore a CCD detector coupled to a telescope\cite{Kus07} improving the signal to background ratio by a factor 150. In 2006 the TPC started to show a degraded performance due to aging. It was then decided to replace the sunset TPC and the existing Micromegas detector in the sunrise side by a new generation of Micromegas detectors\cite{Microbulk,MicrobulkTPC} that coupled with suitable shielding would improve greatly their performance. The new detectors
were commissioning end of 2007 and by mid 2008 they have already shown an improvement in performance that has been translated in a background reduction of a factor 15 compared to the TPC performances and a factor 3 compared to the standard Micromegas detector used without shielding till 2006. The CCD detector will also be upgraded by the fall 2009 by a new detector with improved performance: better low energy response, lower intrinsic background by using more radio-pure materials and less out of time events.

In 2005, the experiment went through a major upgrade to allow operation with He buffer gas in the cold bore. This upgrade was done in two steps: first the system was designed for operation using $^{4}{\rm He}$ and in 2007 the system was upgraded for operation at higher buffer gas densities using $^{3}{\rm He}$. The system has been designed to control the injection of He in the magnet bores with precision and to monitor accurately the gas pressure and temperature\cite{Tapio08,He3TDR}. Special care has been taken to achieve high precision in the reproducibility of steps ($< 0.01\;$mbar) and to protect the system for $^{3}{\rm He}$ loss. The $^{3}{\rm He}$ system has been operating succesfully since december 2007.

\section{Results}

As during phase I, the tracking data (magnet pointing the sun) represented about 2$\times$1.5 hours per day while the rest of the day was used to measure background.  The procedure was to daily increase the $^{4}{\rm He}$ density so that sunrise and sunset detectors measure every pressure. Every specific pressure of the gas allows to test a specific axion mass having a new discovery potential. The $^{4}{\rm He}$ data recorded  end of 2005 and 2006 represents around 300 hours of tracking data and 10 times more hours of background data for each detector, covering 160 pressure settings allowing to scan a new axion mass range between 0.02 and 0.39~eV.

\begin{figure}[hb]
\centerline{\includegraphics[width=0.6\textwidth]{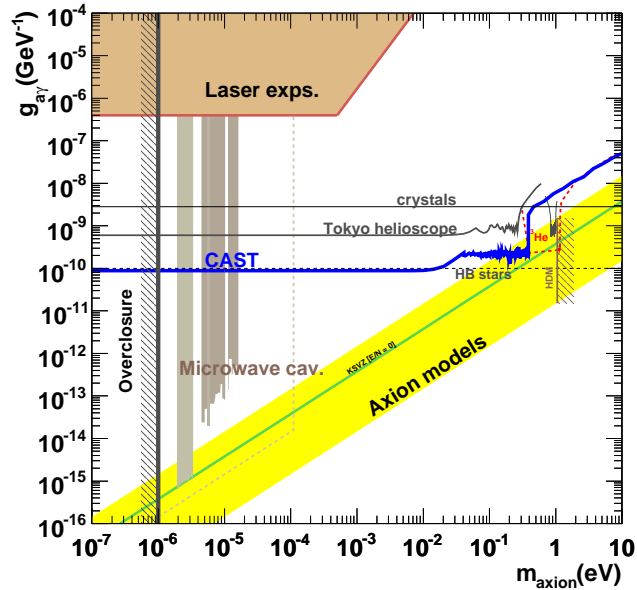}}
\caption{Exclusion plot in the axion-photon coupling versus the axion mass plane for a wide range of parameters. The limit achieved by the CAST experiment (combined result of the CAST phase I
   and $^{4}{\rm He}$ part of phase II) is compared with constraints obtained from the Sumico experiment (the Tokyo helioscope) and HB stars. The red dashed line shows our prospects for the $^{3}{\rm He}$ run started in March 2008. The vertical line (HDM) is the
   hot dark matter limit for hadronic axions
   $\ma<1.0~{\rm eV}$ inferred from observations of the cosmological large-scale structure.
   The yellow band represents typical theoretical models with
   $\left|E/N-1.95\right|$ in the range 0.07--7 where the green solid line
   corresponds to the case when $E/N=0$ is assumed. Limits from laser, microwave and underground detectors~\cite{baudis} for axion searches have been included.}\label{Fig:Exclusion}
\end{figure}

An independent analysis was performed for each data set of the three different detectors. A combined preliminary result was derived  where from the absence of a signal above background CAST excludes a new range in the \gag--\ma~plane shown in figure~\ref{Fig:Exclusion} from axion masses of 0.02\,eV (Phase I) up to masses of 0.39\,eV. This parameter space was not previously explored in laboratory experiments. CAST has therefore entered the QCD axion band for the first time in this range of axion masses, excluding an important portion of the axion parameter space. The final results have been published in \cite{He4CAST}.

The collaboration has performed by-product analysis of the data taken, to look for other
axion scenario to which CAST would also be sensitive. The TPC phase I data has been reanalysed in order look for 14 keV axions coming from M1 transitions. In addition, data taken with a calorimeter during the phase I, 
were used to search for high energy (MeV) lines from high energy axion conversion~\cite{Kresso,Kressodurham}. More recently a few days of data were taken with a visible detector coupled to one end of the CAST magnet~\cite{Marin}, in search for axions with energy in the "visible" range. It is foressen that a permanent setup will be installed in the experiment in order to take data without interfering with the standard program of CAST.

At present, we are running with  $^{3}{\rm He}$ since 2008 and by the end of this year we shoud have reached sensitivities of around $\ma<$0.8\,eV.

\section{After CAST?}

CAST original physics program will finish at the end of 2011. The CAST collaboration is looking into possibilities in order to achieve greater sensitivities. The sensitivity of helioscope axion searches depends
strongly on the magnet's characteristics. Ongoing R \& D on dipoles will lead to stronger and bigger magnets in the coming years (2013-2015). Such magnets could either be adopted to the existing infrastructure of CAST, thus eliminating the cost to the magnet itself, or be used with a new tracking and cryogenics system, improving thus many characteristics (like daily tracking time, safety etc) but increasing the cost. In
parallel, effort will be devoted on the development of high efficiency focusing devices and new electronics for the detectors, aiming to achieve very low background levels. All this would allow pushing the sensitivity of the experiment to the level of 10$^{-11}$GeV$^{-1}$, probing the QCD axion model region for masses higher than 10$^{-2}$\,eV. In order to exceed the limit of 10$^{-11}$GeV$^{-1}$  with a helioscope, specially developed magnets are necessary. Taking advantage of the fact that the important parameter is the strength and not the homogeneity of the magnetic field, a stronger magnet with bigger aperture, could be designed and constructed especially for such an experiment. This is a more expensive and long term option, which would allow reaching the limits of the helioscope axion searches with an increased discovery potential. These different options as well as expected sensitivities were presented in detail in~\cite{futureCAST}.


\section{Conclusions}

The CAST experiment has established the most stringent experimental limit on axion coupling constant over a wide range of masses, exceeding astrophysical constraints.  The $^{4}{\rm He}$ phase has allowed to enter in an unexplored region favoured by the theory axion models. From the absence of excess X-rays when the magnet was pointing to the Sun, we set a preliminary upper limit on the axion-photon coupling of $\gag\lesssim 2.22\times 10^{-10}\,{\rm GeV}^{-1}$ at 95\% CL for $\ma \lesssim 0.4$~eV, the exact result depending on the pressure setting. At present, with the $^{3}{\rm He}$ run we are exploring deeper this region to reach sensitivities of $\ma<$1.2\,eV.
The Collaboration is looking into developping the new generation of helioscopes in order to reach sensitivities of the order of 10$^{-11}$GeV$^{-1}$ leading to explore a large part of the QCD favoured model region including the otherwise non-accessible sub-keV range.



\begin{footnotesize}



%

\end{footnotesize}


\end{document}